\documentclass[]{spie}  

\usepackage{amsmath,amsfonts,amssymb}
\usepackage{graphicx}
\usepackage[colorlinks=true, allcolors=blue]{hyperref}
\usepackage{authblk}

\title{Mechanical design and development of TES bolometer detector arrays for the Advanced ACTPol experiment}

\author[a]{Jonathan T. Ward}
\author[b]{Jason Austermann}
\author[b]{James A. Beall}
\author[e]{Steve K. Choi}
\author[e]{Kevin T. Crowley}
\author[a]{Mark J. Devlin}
\author[b]{Shannon M. Duff}
\author[c]{Patricio M. Gallardo}
\author[c]{Shawn W. Henderson}
\author[e]{Shuay-Pwu Patty Ho}
\author[b]{Gene Hilton}
\author[b]{Johannes Hubmayr}
\author[a]{Niloufar Khavari}
\author[a]{Jeffrey Klein}
\author[c]{Brian J. Koopman}
\author[d]{Dale Li}
\author[i]{Jeffrey McMahon}
\author[a]{Grace Mumby}
\author[a]{Federico Nati}
\author[c]{Michael D. Niemack}
\author[e]{Lyman A. Page}
\author[e]{Maria Salatino}
\author[f]{Alessandro Schillaci}
\author[a]{Benjamin L. Schmitt}
\author[e]{Sara M. Simon}
\author[e]{Suzanne T. Staggs}
\author[g]{Robert Thornton}
\author[b]{Joel N. Ullom}
\author[c]{Eve M. Vavagiakis}
\author[h]{Edward J. Wollack}

\affil[a]{Department of Physics and Astronomy, University of Pennsylania, 209 S 33rd Street, Philadelphia, PA, USA 19104}
\affil[b]{NIST Quantum Devices Group, 325 Broadway Mailcode 817.03, Boulder, CO, USA 80305 }
\affil[c]{Department of Physics, Cornell University, Ithaca, NY, USA 14853}
\affil[d]{SLAC National Accelerator Laboratory, 2575 Sandy Hill Road, Menlo Park, CA 94025}
\affil[e]{Joseph Henry Laboratories of Physics, Jadwin Hall, Princeton University, Princeton, NJ, USA 08544}
\affil[f]{Institute of Astrophysics, Pontificia Universidad Catolica de Chile, Avda. Libertador Bernardo O'Higgins 340, Santiao, Chile}
\affil[g]{Department of Physics, West Chester University of Pennsylvania, 700 South High Street, West Chester, PA, USA 19383} 
\affil[h]{NASA/Goddard Space Flight Center, Observational Cosmology Laboratory, 8800 Greenbelt Rd, Greenbelt, MD 20771, USA}
\affil[i]{Department of Physics, University of Michigan Ann Arbor, Randall Labs, 450 Church Street, Ann Arbor, MI 48103, USA}

\begin{document} 
\maketitle

\begin{abstract}
The next generation Advanced ACTPol (AdvACT) experiment is currently underway and will consist of four Transition Edge Sensor (TES) bolometer arrays, with three operating together, totaling $\sim$5800 detectors on the sky.  Building on experience gained with the ACTPol detector arrays, AdvACT will utilize various new technologies, including 150~mm detector wafers equipped with multichroic pixels, allowing for a more densely packed focal plane.  Each set of detectors includes a feedhorn array of stacked silicon wafers which form a spline profile leading to each pixel.  This is then followed by a waveguide interface plate, detector wafer, back short cavity plate, and backshort cap.  Each array is housed in a custom designed structure manufactured from high purity copper and then gold plated.  In addition to the detector array assembly, the array package also encloses cryogenic readout electronics.  We present the full mechanical design of the AdvACT high frequency (HF) detector array package along with a detailed look at the detector array stack assemblies.  This experiment will also make use of extensive hardware and software previously developed for ACT, which will be modified to incorporate the new AdvACT instruments.  Therefore, we discuss the integration of all AdvACT arrays with pre-existing ACTPol infrastructure.  
\end{abstract}


\section{INTRODUCTION}
The Atacama Cosmology Telescope (ACT) \cite{ACT} is a six-meter diameter telescope located at an elevation of 5,190 meters on Cerro Toco in the Andes mountains of northern Chile.    With a broader frequency range and higher sensitivity than the first polarimeter used for ACT (ACTPol\cite{BobInstrument}), AdvACT aims to make high resolution measurements of the polarized Cosmic Microwave Background (CMB) radiation over a range of angular scales.  These observations will probe properties of the universe such as the tensor to scalar ratio $r$ and the sum of the neutrino masses, while also further constraining standard $\Lambda$CDM model parameters (e.g. Spergel et al. 2003\cite{Spergel} and Sievers, Hlozek, Nolta et al. 2013\cite{Sievers}).  In order to achieve these and other science goals, the AdvACT instrument \cite{HendersonLTD} will consist of four multichroic, polarization-sensitive TES bolometer detector arrays\cite{LiLTD}.  Each array will be sensitive to two frequency bands:  150 and 230 GHz in a high frequency (HF) array, 90 and 150 GHz in two middle frequency (MF) arrays, and 28 and 41 GHz in a low frequency (LF) array.  The detectors are cryogenically cooled to 100~mK using a dilution refrigerator (DR), which allows for continuous cooling, increases detector sensitivity, and is also required for many of the superconducting components of the detectors and readout to operate.  This paper discusses the design, fabrication and assembly of the AdvACT detector and feedhorn wafers as well as the mechanical array structure used to house all of the 100~mK electronic components.  We also describe new structural components that have been built to allow AdvACT array packages to integrate with the existing ACTPol receiver.

\section{FEEDHORN-COUPLED DETECTOR STACK ASSEMBLY}
 AdvACT will feature multichroic technologies that allow a single pixel to simultaneously measure two separate frequencies\cite{Datta}.  The AdvACT focal plane consists of a four-piece detector stack coupled to a wideband spline-profiled feedhorn array.  The wafers are fabricated at the National Institute of Standards and Technology (NIST) by ACT collaborators in the Quantum Sensors Group\cite{NIST}.  AdvACT has successfully produced science-grade 150mm multichroic detector wafers for CMB experiment, allowing for an increase in the total number of detectors on the sky.  The features of each array, such as feedhorn aperture and profile, pixel spacing, and pixel design, will differ slightly depending on the operating frequency, but the overall mechanical design concepts will remain consistent for all of the AdvACT arrays.  Here, we present the feedhorn and detector stack designs for the HF array, which is currently being assembled and tested and will be deployed in June 2016.  The MF and LF arrays are still in the design phase.

\subsection{AdvACT High Frequency (HF) Detector Stack}
\begin{figure}[t] 
   \centering
   \includegraphics[width=6.75in]{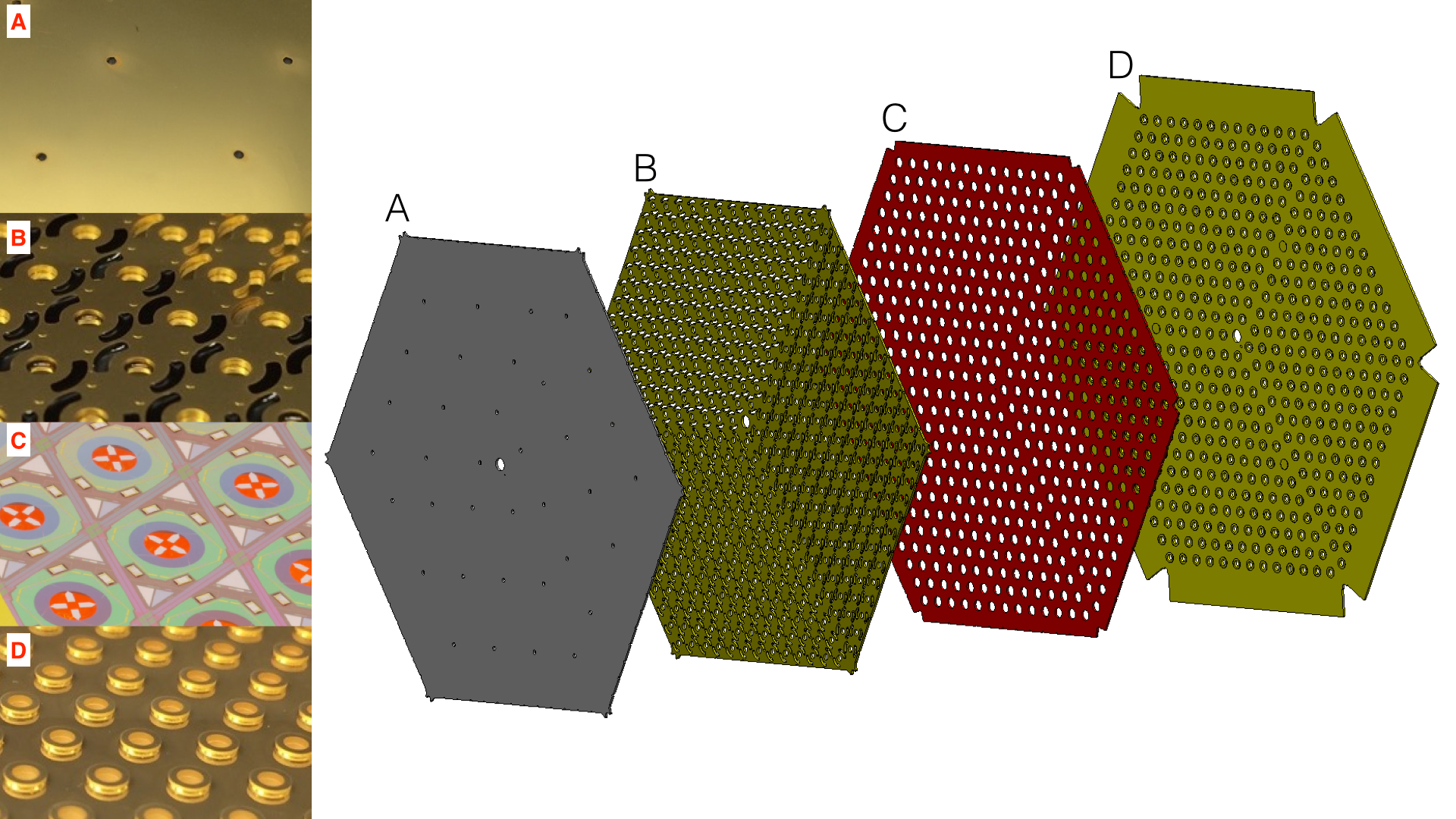} 
   \caption{Zoomed and exploded views of the AdvACT HF detector stack wafers.  (A).  In addition to serving as the back short of the detector stack, the backshort cap wafer also mechanically protects and heat sinks the wafers in the detector stack.  (B).  The backshort cavity wafer opens the back of the OMT on the detector wafer to the reflective back short cap, providing a quarter-wave cavity.  The semi-circular cutouts around it's openings (called ``moats'') are filled with absorptive CR-110 to absorb excess radiation and minimize light leakage between pixels.  (C).  The detector wafer houses the individual pixels and detector wiring.  (D) The waveguide interface plate serves as a seal between the feedhorns and detector wafer and uses boss features on the detector side of the wafer to prevent photon loss.}
   \label{DetStack}
\end{figure}

The AdvACT HF detector stack consists of four individual silicon wafers.  The detection scheme is based on placing polarization sensitive probes inside a circular waveguide, and the four-part silicon detector stack realizes this scheme in arrays of detectors.  A model of each wafer can be seen in Figure \ref{DetStack}.  The outline and inner features of each wafer are created by etching the silicon using deep reactive ion etching\cite{ShannonLTD}.  Photoresist masks of the mechanical features are imprinted on the wafers, leaving the silicon wafer exposed where material is then removed by the etching.  After etching, a seed layer is added to all wafers but the detector wafer, which are then electroplated with a layer of gold.  The four wafers are stacked at the back side of the feedhorn array and held in place by a spring loaded structure, which will be discussed later in Section 3.3.

The wafer closest to the feedhorns is the waveguide interface plate (WIP).  The WIP is 500~$\mu$m thick and serves as an interface between the feedhorn array and detector wafer.  Small boss features on the back of the WIP mate with the openings in the detector wafer to limit the loss of photons at the feedhorn/detector interface.  The bosses act as an extension of the feedhorn array and guide the photons to the detector probes.  A magnified view of the WIP features can be seen in box D of Figure \ref{DetStack}.

The next component in the stack is the detector wafer.  The detector wafer is also 500~$\mu$m thick and is populated with the AdvACT pixels.  Each pixel is an integrated superconducting circuit that separates linear polarization, defines the observation passband, and senses radiation with TES bolometers.  A detailed view of several AdvACT pixels\cite{Datta} on the detector wafer is shown in box C of Figure \ref{DetStack}.  The detector wafer also contains traces that couple each detector probe's signal to the readout electronics.

The third and fourth wafers together provide quarter wave backshorts behind the detectors to improve the bolometers' efficiency.  The third wafer is the backshort cavity plate.  The cavity plate features openings to the backshort cap at the center of each pixel, providing a reflective surface behind the OMT and defining the cavity length.  Additionally, around each opening is a semi-circular ``moat'' cutout behind the TES detectors that is filled with an absorptive material (CR-110) and are shown in box B of Figure \ref{DetStack}.  Radiation leakage from the waveguide can result in crosstalk between individual TESes, including leakage to orthogonal polarization modes within the same pixel.  The moats aim to absorb the excess radiation and limit these effects.  Finally, the stack finishes with the backshort cap.  In addition to acting as the reflective back surface of the quarter-wave cavity, the backshort cap provides mechanical protection for the detector stack and serves as a heat sink.  Gold ribbons are gap-welded to the backshort cap and then connected to the detector ring (discussed in Section 3.3) of the array package.  Holes in the back of the backshort cap are used for gluing the cap to the backshort cavity wafer.  The gold-plated surface of the backshort cap is shown in box A of Figure \ref{DetStack}.

The four wafers are aligned with pins and glued together by filling semi-circular corrals along the edges of the detector wafer, backshort cavity plate, and backshort cap with epoxy.  The stack is then integrated with the feedhorn array, readout electronics, and array package structure.

\begin{figure}[t] 
   \centering
   \includegraphics[width=6.75in]{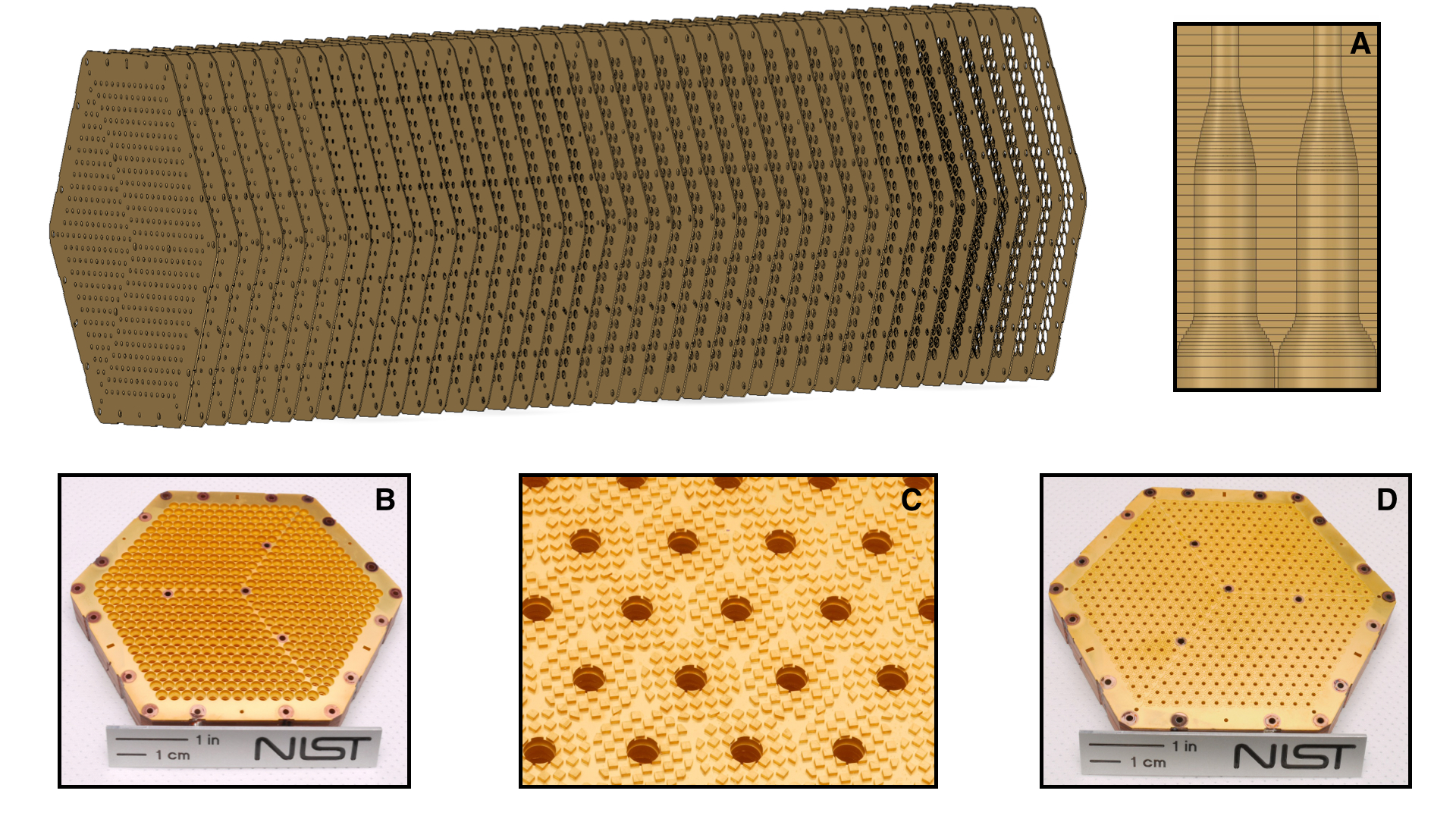} 
   \caption{Exploded view of the AdvACT HF feedhorn stack model.  The full feedhorn array consists of 41 individual silicon wafers.  The wafer thicknesses vary from 333~$\mu$m to 500~$\mu$m depending on where in the waveguide profile they are located.  Each wafer exhibits different sized apertures resulting in a spline-profiled waveguide when fully assembled.  (A) Cross section view of the feedhorn profile created by the stack, with the bottom being the sky side and top being the detector side.  Also note the double etching in the smaller 333~$\mu$m wafers.  (B) The sky side of the AdvACT HF feedhorn array.  This is where the aperture is the largest.  (C) A close-up view of the photonic choke features on the detector side of the HF feedhorn array.  (D) The detector side of the HF feedhorn array.  This is were the aperture is the smallest.  The detector stack is aligned on the surface shown and the WIP described in Section 2.1 guides photons from the waveguide to to the detector probes.}
   \label{Feedhorns}
\end{figure}

\subsection{AdvACT High Frequency (HF) Feedhorn Stack}
The feedhorn arrays for the AdvACT detectors are made by stacking individual silicon wafers with different sized pixel apertures, forming a spline-profiled, smooth-walled waveguide.  A cross section of the waveguide profile is shown in box A of Figure \ref{Feedhorns}.  More details on its design, including the photonic choke mentioned below, are found in Simon et al\cite{SimonSPIE}.  Each wafer is cut using deep reactive ion etching.  After etching, the wafers are cleaned in Nano-strip\textsuperscript{\textregistered} to remove any organic material, rinsed in deionized water, and spun dry.  After cleaning, each wafer is sputter coated with a seed later of 200~nm Ti and 1~$\mu$m Cu.

Next, the wafers are stacked and aligned using the flat edges of the wafers along with several alignment pin holes.  Once alignment is completed, the stack is clamped using perimeter and center screws with nylon washers.  The outer edges of the wafers are equipped with indentations where Stycast\textsuperscript{\textregistered} 2850 is used to glue the wafers together.  In preparation for electroplating, copper braids are glued in several locations for electrical contact.  The full stack is then electroplated with 3~$\mu$m of copper followed by 3~$\mu$m of gold.  Figure \ref{Feedhorns} shows an exploded view of the HF feedhorn stack model and several views of the completed HF feedhorn array.  The small boss features shown in box B of Figure \ref{Feedhorns} create a photonic choke on the detector side of the final wafer in the feedhorn array.  The photonic choke substantially reduces leakage between the feedhorn array and WIP\cite{Wollack}.  

The detector stack described in Section 2.1 is wire bonded to a PCB housing biasing and multiplexing components\cite{HendersonLTD} and then aligned on the back of the waveguide using alignment pins and held in place, as described in Section 3.3, rather than glued.  See Li et al.\cite{Yaq} for more details on the detector assembly process.  Finally, the full detector/feedhorn assembly is suspended at the focal plane of the optics chain in the array package structure using the hardware described in Section 3.3.

\section{ARRAY PACKAGE STRUCTURE}
All AdvACT detector and feedhorn arrays are supported by and contained within the array package.  The array package protects the delicate camera elements, heat sinks these elements to 100~mK, places the array at the focus of the optics, and serves as a structural support for both the arrays and readout electronics.  Each of the four individual components of the array package are manufactured from Oxygen-Free High-Conductivity (OFHC) copper and gold plated to maximize thermal conductivity.  The design and purpose of each array package element is described below.  The full assembly is built and tested before being deployed in the receiver as a single unit.  

\begin{figure}[t] 
   \centering
   \includegraphics[width=6.75in]{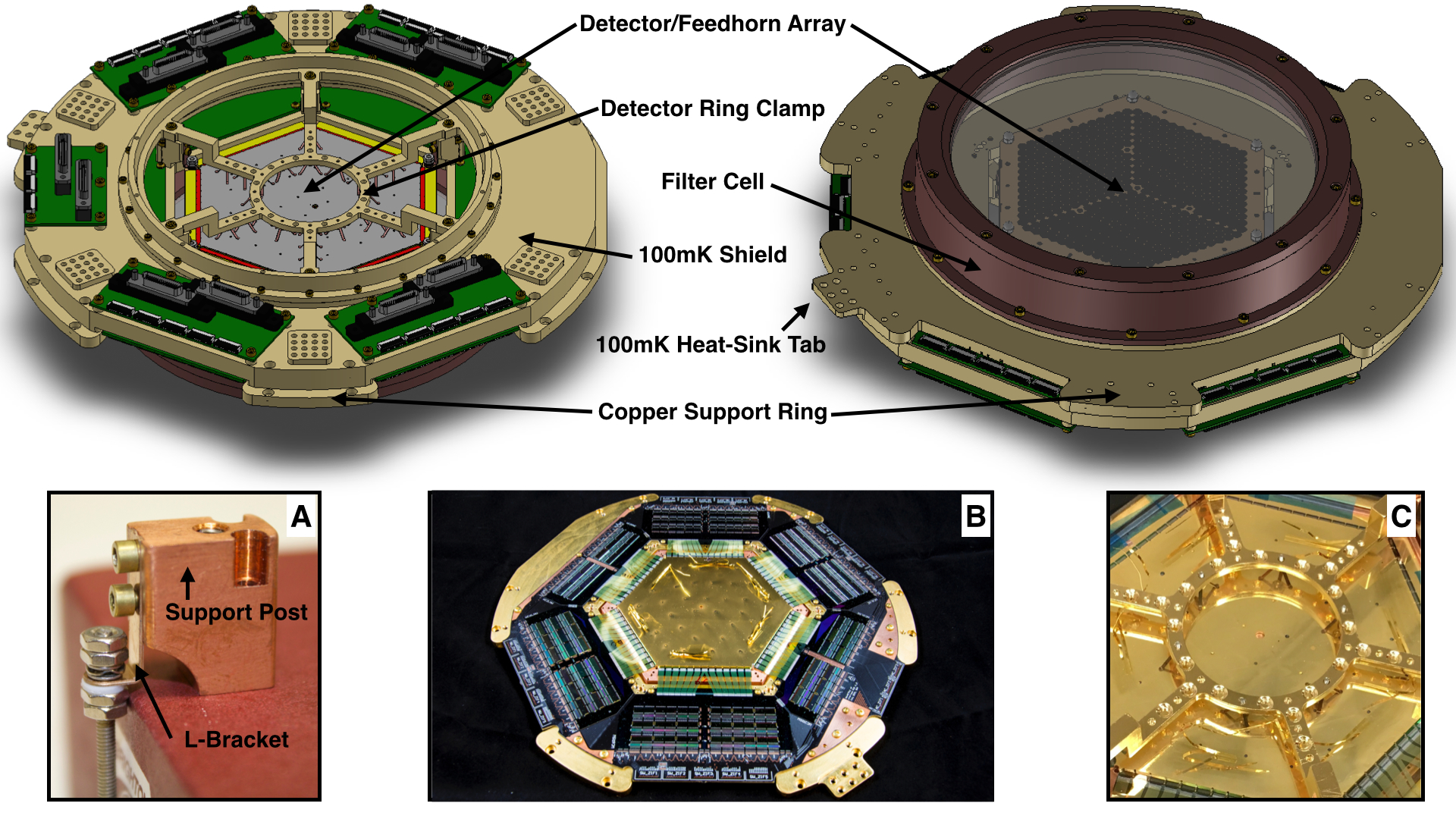} 
   \caption{The full AdvACT HF array package.  Top left shows a drawing of the backside view and the top right shows a drawing of the sky-side view.  The backside consists of the main PCB mounted to the copper support ring and covered by the 100~mK shield.  The lid of the shield is removed in the model to show the array suspension and detector ring clamp.  Five interface boards are mounted to the top surface of the 100~mK shield and connect the 100~mK detector electronics to the 1K series arrays.  The sky-side of the array package shows the entrance aperture of the feedhorn array enclosed by a cell that houses a low-pass filter.  (A)  A close up view of the hardware used to suspend the detector/feedhorn array in the array package.  A total of six threaded stainless steel rods penetrating the length of the feedhorn array are secured to copper-beryllium L-brackets and bolted to the gold-plated copper support posts.  (B)  The AdvACT HF array with 100~mK shield and detector ring clamp removed during the electrical assembly process.  The main PCB, populated with multiplexing and bias chips for readout, surrounds the detector array and is attached to the detectors via flexible circuitry and wire bonds.  (C)  A zoomed view of the detector ring pressing against the back of the detector stack.  Copper-beryllium tripod springs on the bottom surface of the detector ring clamp keep pressure on the back of the stack for mechanical support.  Gold ribbon is gap-welded to the backshort cap in the detector stack and screwed onto the detector ring clamp for heat sinking.}
   \label{ArrPack}
\end{figure}

\subsection{Copper Support Ring}
The copper support ring is used as the mounting surface for the 100~mK readout PCB and also serves as the main structural support of the entire array package.  A monolithic flat PCB, populated with multiplexing and bias chips for readout, is mounted directly to the top surface of the ring, where it is then wire-bonded to the detector array via flexible superconducting circuitry\cite{PappasLTD}.  In ACTPol, nine separate boards were wire bonded to the detector array and turned 90 degrees to vertical due to radial space constraints.  The new AdvACT design keeps the PCB and flex securely fastened during the entirety of the assembly process, minimizing the chances of flex or wire bond damage.  Additionally, small trench patterns are etched into the surface of the support ring to match the via locations on the bottom of the PCB, preventing shorts.

A small tab on the perimeter of the ring, shown in Figure \ref{ArrPack}, acts as the 100~mK connection for heat straps connected to the DR.  By attaching the array package directly to the mixing chamber plate of the DR, we reduce the total number of 100~mK thermal joints by one when compared to ACTPol, which should improve cryogenic performance.  Mounting holes on the inner diameter are used to attach the support spokes that suspend the detector/feedhorn array and also to secure the detector ring in place.  Mounting holes on the outer diameter are used to connect the 100~mK shield to the assembly and to mount the entire array package in the receiver.

\subsection{100mK Shield}
The 100~mK shield of the array package attaches directly to the top of the copper support ring after the main PCB is mounted.  Figure \ref{ArrPack} shows the shield with the lid removed.  When the lid is in place, the shield completely seals off the electronics and detector/feedhorn array from the outside.  This protects the delicate components of the camera during testing and transport while also providing a light-tight cavity when the array package is installed in the receiver.  The 100~mK shield is also used as a mounting surface for the five interface boards shown in the top left of Figure \ref{ArrPack}.  These boards provide the interface between the 100~mK main PCB and the 1K series array readout electronics.

\subsection{Support Posts and Detector Ring}
The support posts and detector ring are the main support structures for the detector/feedhorn stack.  Six gold-plated posts mount directly to the inner diameter of the copper ring at each vertex of the hexagonal cutout in the copper support ring and PCB, and couple with copper-beryllium L-brackets to suspend the detector/feedhorn array in the focal plane.  A detailed look at the hardware used to secure the feedhorn array to the L-brackets and posts can be seen in box A of Figure \ref{ArrPack}.

Once the array is in place, the detector ring clamp bolts to the top of the posts and holds the detector stack in place on top of the feedhorn array.  The bottom edge of the detector ring clamp is equipped with copper-beryllium tripod springs that press on the back of the detector stack, serving as both a heat sink and mechanical support.  Similar methods were used in ACTPol, but six separate fixtures were used for support.  Transition to a monolithic design ensures even pressure across the entire array and allows for a more simple and safe assembly process.  Box C of Figure \ref{ArrPack} shows the detector ring pressed against the back of the HF detector stack during assembly and testing.

\subsection{Filter Cell}
A filter cell is mounted to the bottom of the copper support ring and houses a low-pass filter placed in front of the sky-side aperture of the feedhorn array.  The filter cell is composed of two parts; the body and the clamp and matches the design used in ACTPol\cite{BobInstrument}.  A groove is cut along the perimeter of the body and contains an electromagnetic interference gasket.  The gasket is a flexible spiral wound from spring temper beryllium copper and tin plated\cite{SPIRA}.  This makes for exceptional spring memory and compression as well as excellent conductivity and shielding properties.  The filter is then placed on top of the gasket and compressed down using the clamp, resulting in a tight hold even after thermal contraction at low temperatures.

\begin{figure}[t] 
   \centering
   \includegraphics[width=6.75in]{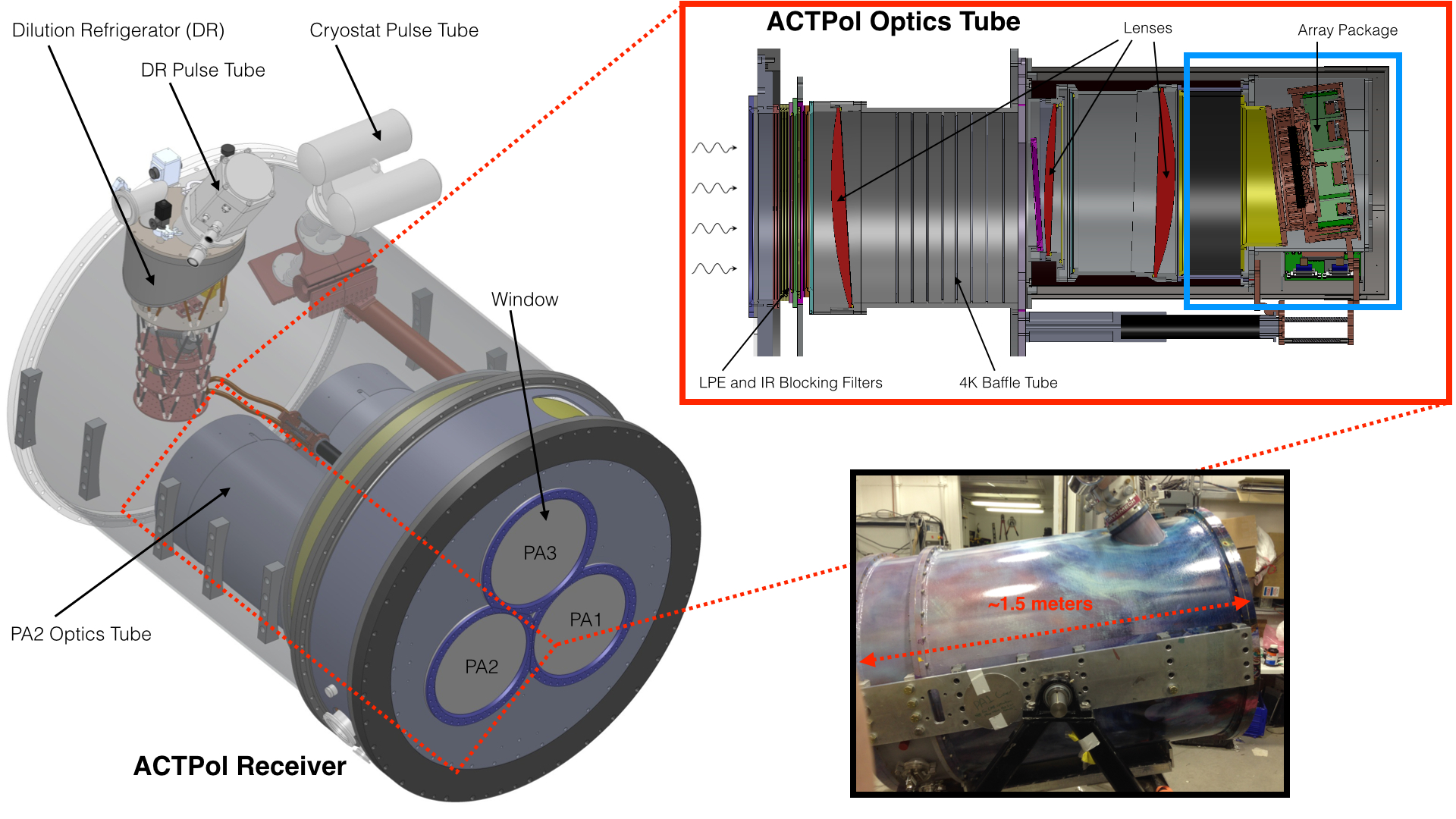} 
   \caption{A general overview of the ACTPol receiver.  The left side of the figure shows the full cryostat model, including the DR and pulse tube inserts.  There are a total of three separate optics tubes, each of which contains a separate detector array package.  PA1, PA2, and PA3 are references to the ACTPol detector arrays that were deployed in each optics tube and will be replaced by the LF, MF, and HF AdvACT detector arrays.  The red box shows a cross section of one of the optics tubes.  Each tube contains several temperature stages, as well as various filters and lenses that make up the re-imaging optics chain.  The detectors sit inside the array package at the back of the optics tube at the focal plane of the optics design.  The blue box indicates the region discussed in Figure 5, where the new AdvACT receiver components and detector arrays will be installed.  The bottom right shows a photograph of the fully assembled ACTPol receiver.  }
   \label{ACT_Receiver}
\end{figure}

\section{Additional Receiver Modifications}
The AdvACT array packages will be deployed in the cryogenic receiver built for the previous generation ACTPol experiment\cite{BobInstrument}.  A brief overview of the ACTPol receiver is shown and discussed in Figure \ref{ACT_Receiver}.  Most of the infrastructure from ACTPol will be reused, but due to significant changes in the array package size and shape, a majority of the 1K structural components have been redesigned.  This includes the main support structure for the array package, the radiation shield cavity surrounding the array, the series array electronics support structure, and optics tube lids responsible for shielding the array from magnetic fields.  The section of the ACTPol cryostat receiving upgrades is highlighted in blue in Figure \ref{ACT_Receiver} and further detailed in Figure \ref{HF_Tube}.

\subsection{Array Support Wedge}
The array package is mounted to an angled garolite (G10) ``wedge'' surface inside the optics tube.  The G10 wedge and array package are the only components of the receiver at 100mK and are thermally isolated by mounting directly to a carbon fiber suspension.  The wedge is manufactured from G10 to further promote thermal isolation from the warmer optics tube components.  The wedge angle and orientation align the feedhorn array with the telecentric image of the sky produced by the reimaging optics\cite{BobInstrument}.  The mounting surface for the array package matches the outline of the copper support ring, providing a stable and secure hold for the AdvACT camera.  The wedge is labeled in the cross section model shown in Figure \ref{HF_Tube} and the final manufactured part is shown in box A of Figure \ref{HF_Tube}.

\subsection{1K Radiation Shield}
\begin{figure}[t] 
   \centering
   \includegraphics[width=6.75in]{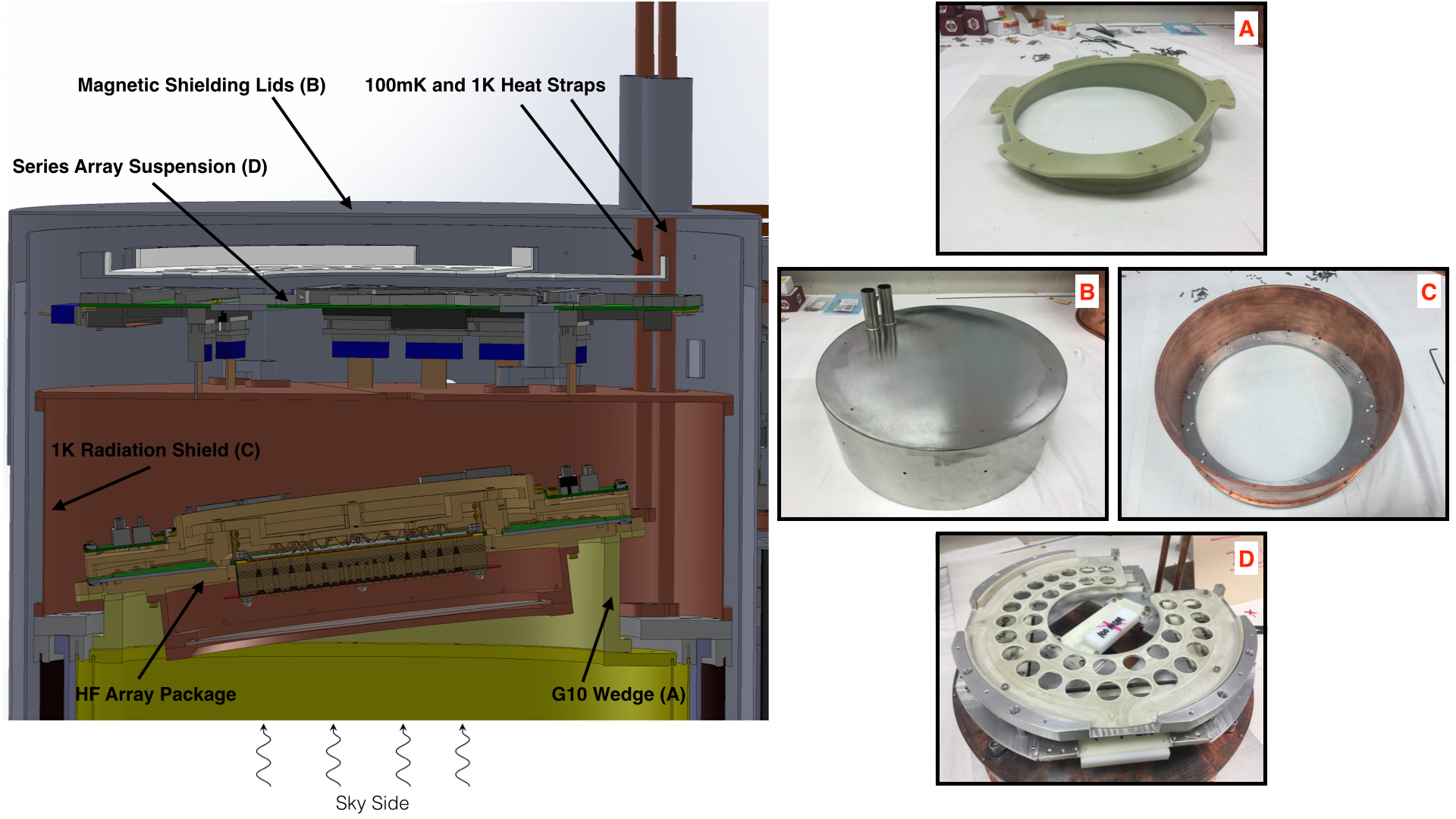} 
   \caption{A cross section view of the lower HF optics tube.  The individual components designed for the HF deployment are labeled in black.  (A)  The G10 wedge is used to structurally support the HF array package while also helping thermally isolate the detectors from warmer stages of the receiver.  (B)  Each optics tube is surrounded by cryoperm magnetic shielding.  The lids for AdvACT have been redesigned to have openings for the new 1K and 100mK heat straps.  (C)  The total array package is larger in diameter and flatter when compared to ACTPol, leading to a necessary reformulation of the 1K radiation shield.  The shield protects the array package and also serves as a 1K surface to which the series array suspension is attached.  (D)  Here, plastic mechanical-model PCBs mounted to the aluminum series array suspension are visible through the holes in the cable routing plate.   The routing plate is placed above the electronics to properly secure the Tekdata\textsuperscript{\textregistered} cables used for detector readout.  The entire structure is attached to the top of the 1K radiation shield.}
   \label{HF_Tube}
\end{figure}

 The 1K radiation shield, shown in the cross section and box C of Figure \ref{HF_Tube}, is an OFHC copper cavity that encloses the entire array package and also serves as a 1K thermal connection.  The radiation shield is secured directly to the 1K main plate, which is connected to the 1K stage of the DR using 3/8~inch 4N9 copper rods as heat straps (shown on the right side of Figure \ref{HF_Tube}).  Furthermore, the support structure for the series array readout electronics mounts to the lid of the 1K radiation shield.  Tekdata Cryoconnect\cite{TekData} cables connecting the array package to the series array electronics are threaded through an opening in the center of the lid.  The series array structure also uses the 1K radiation shield as a thermal contact to heat sink the electronic components to 1K.
 
 \subsection{Series Array Suspension}
 The HF series array electronics are held in place and heat sunk by a 6061 aluminum structure mounted to the top of the 1K radiation shield lid.  The series array suspension is composed of two primary pieces, the electronics board mounting plate and the cable routing plate.  A total of five PCBs are bolted to the mounting plate, with connectors going to the array pointed toward the detector assembly and connectors leaving the optics tube pointing out radially.  
 
The cable routing plate is placed directly above the mounting plate and keeps the cables from laying directly on the PCBs.  It also allows for direct routing of the cables leaving the optics tube through gaps in the cryoperm magnetic shielding described in Section 4.4.  Improved handling of the cables was motivated by a small number of critical line failures in the ACTPol cryogenic wiring, which could have been a result of cable mismanagement.  The top surface of the cable route plate (where cables will be tied down) is also lined with G10 to reduce the risk of shorts between the TekData\textsuperscript{\textregistered} cables and the aluminum structure.  The fully assembled series array suspension populated with plastic mock series array PCBs is shown in box D of Figure \ref{HF_Tube}.  
 
 \subsection{Magnetic Shielding Lids}
 The HF optics tube is surrounded by two nested cryoperm cylinders for magnetic shielding as described in Thornton et al\cite{BobInstrument}.  Cryoperm is a cryogenic, high permeability magnetic shielding combination of nickel, copper, molybdenum and iron\cite{Amuneal}.  External magnetic fields can interfere with the detectors and the SQUIDs used for multiplexing, so minimizing these effects is essential.  However, under the new AdvACT design, the 100~mK and 1K thermal connections between the DR and interior optics tube surfaces go through the back of the optics tube rather than the side.  For AdvACT, the lids of the magnetic shields have been designed to accommodate this change.  The outer cryoperm lid includes tubular ``snouts'' to reduce the penetration of magnetic fields.  These features can be seen in box B of Figure \ref{HF_Tube}.

\section{Future Work}
The AdvACT experiment will continue with further camera/receiver upgrades and CMB observations for the next three years.  In addition to the HF array discussed in this paper, two MF arrays and an LF array will be designed, built, and deployed on the telescope in Chile.  A majority of the mechanical designs for the MF and LF arrays will mirror what has been presented for the HF array.  Changes to the feedhorn and detector stack design will optimize the arrays for different frequency bands.  Similar changes will also be made to the receiver infrastructure for both the MF and LF optics tubes, including a transition to more direct heat sinking for all three arrays.  When completed, AdvACT will observe nearly half the sky with approximately 5800 detectors and cover five frequency bands.  These increases in sky coverage, sensitivity, and frequency coverage will continue to push the Atacama Cosmology Telescope to the forefront of modern experimental cosmology.  

\acknowledgements
The Advanced ACT Experiment is supported by the U.S. National Science Foundation through award 1440226.  The work presented in this paper is supported by a NASA Space Technology Research Fellowship grant number NNX15AQ74H titled ``Development of Inflation Probe Technologies for the Advanced ACT Experiment''.  The development of multichroic detectors and lenses was supported by NASA grants NNX13AE56G and NNX14AB58G. The work of KPC, KTC, EG, BJK, CM, BLS, and SMS was supported by NASA Space Technology Research Fellowship awards.

\end{document}